# Separability criterion for *n*-particle states

R. hamzehofi

Department of Physics, Faculty of Science, Shahid Chamran University of Ahvaz, Ahvaz, Iran

**Abstract**

This research introduces the concept of the purity number, which represents the number of separable *s*-particle sub-states within an *n*-particle state ($s < n$). It establishes that, for any $s$, achieving the maximum purity number is both a necessary and sufficient condition for the separability of *n*-particle pure states, and a necessary condition for the separability of *n*-particle mixed states. Subsequently, the study delves into the concept of entanglement rate in *n*-particle pure states. The entanglement rate of an *n*-particle pure state, in which all entangled sub-states are maximally entangled, can be considered as a measure of entanglement.

## 1. Introduction

In the quantum information theory, two fundamental concepts reign supreme: separability and entanglement. The concept of separability, which dictates the extent to which the states of individual particles within a composite system can be understood independently of one another. In essence, separable states allow us to treat each particle as its own entity, devoid of any entangling influences from its counterparts. Entanglement occurs when the quantum states of two or more particles become intertwined in a manner that transcends classical causality, forging connections that persist across vast distances and instantaneously influence one another's behavior. In an entangled state, the fate of each particle is inextricably linked to that of its entangled counterparts, leading to correlations that defy classic explanations. Entanglement is used in various topics such as quantum teleportation, cryptography and superdense coding [1-5]. Unlocking the potential of quantum technologies hinges on revealing the intricate phenomenon of entanglement within quantum states. Yet, detecting this phenomenon is far from straightforward. Identifying the separability of certain states poses a formidable challenge, often classified as NP-hard [6]. Nevertheless, physicists have devised diverse entanglement criteria, each offering unique insights alongside its inherent constraints [7-15].

Some quantum states can be represented as products of multiple sub-states. It becomes apparent that when none of these sub-states are entangled, the overall state is deemed separable. Several inquiries emerge regarding such states. For example, how can one ascertain the separability of all sub-states? Conversely, if there exist entangled sub-states, what is the rate of entanglement for these states? This research endeavors to tackle these inquiries. Hence, the subsequent section introduces the concept of the purity number for *n*-particle states. Leveraging this concept, a criterion is established to identify the separability of *n*-particle states. Furthermore, a method is proposed to quantify the rate of entanglement within an *n*-particle pure state.

E-mail address: rezahamzehofi@gmail.com

https://orcid.org/0009-0002-0036-7358

## 2. Purity number and entanglement rate

### 2.1. Purity

The purity of a quantum state $\rho$, describes how close the state is to being a pure state and is defined as follows [16]:

$$P = tr(\rho^2). \tag{1}$$

The purity satisfies $d^{-1} \leq P \leq 1$, where $d$ is the dimension of the Hilbert space. For $P=1$, the state is pure, while lower values indicate mixed states. Also, $P = d^{-1}$ indicates a maximally mixed state

### 2.2 Purity number as a separability criterion

An $n$-particle pure state ($n \geq 3$) can be decomposed as follows:

$$|\psi\rangle = |\eta_1\rangle \otimes ... \otimes |\eta_m\rangle, \tag{2}$$

where $m \leq n$. All of the sub-states are pure, that is:

$$P(|\eta_1\rangle) = ... = P(|\eta_m\rangle) = 1. \tag{3}$$

Within this study, the "purity number," denoted by $\gamma_s$, characterizes the count of seperable $s$-particle sub-states within an $n$-particle state. Utilizing equations (2) and (3), the purity number of an $s$-particle sub-state is derived by evaluating the purity across all feasible $s$-particle sub-states. If a sub-state's purity equals unity, it contributes to the count; otherwise, it is disregarded. Notably, to compute the density matrix of a sub-state, we employ the below equation:

$$\eta = \sum_i \langle i|_{\bar{\eta}} (|\psi\rangle\langle\psi|) |i\rangle_{\bar{\eta}}, \tag{4}$$

where $\bar{\eta}$ is the complementary set of $\eta$. Moreover, the upper bound of the purity number for an $s$-particle sub-state can be obtained as follows:

$$\gamma_s^{\max} = \binom{n}{s} = \frac{n!}{(n-s)!s!}. \tag{5}$$

Now, consider an entangled sub-state $|\eta_{\alpha\beta}\rangle$ within an $n$-particle pure state $|\psi\rangle$ in arbitrary dimensions as follows:

$$|\psi\rangle = \sum_{i_1,i_2,...,i_n=0}^{d_1-1,d_2-1,...,d_n-1} k_{i_1 i_2 ... i_n} |i_1\rangle \otimes |i_2\rangle \otimes ... \otimes |i_n\rangle, \tag{6}$$

where $\sum k_{i_1 i_2 ... i_n} \cdot k^*_{i_1 i_2 ... i_n} = 1$. In such instances, the purity of multi-particle sub-states incorporating solely one of the entangled particles $\alpha$ or $\beta$ is not unity, i.e.:

$$0 \leq \gamma_s < \gamma_s^{\max}. \tag{7}$$

Essentially, the maximum purity number functions as a separability criterion. An $n$-particle pure state is deemed separable if its purity number reaches its maximum, a condition both necessary and sufficient. Additionally, from this criterion, it follows that if Eq. (7) holds, at least one of the sub-states is entangled. Furthermore, the necessary and sufficient condition for all particles to be entangled together is as follows:

$$\sum_{s=1}^{n-1} \gamma_s = 0. \tag{8}$$

The above equation means the overall state can not be written as product of sub-states. That is, all particles are entangled together.

Now we extend this criterion to *n*-particle mixed states. Generally, an *n*-particle mixed state $\rho$ ($n \geq 3$) can be decomposed into its sub-states in the following form:

$$\rho = \rho_1 \otimes ... \otimes \rho_m, \tag{9}$$

where $1 \leq m \leq n$ and $\rho_i$ represents the density matrix of the *i*th sub-state. Additionally, at least one of the sub-states is mixed. If we denote the number of particles present in pure sub-states as $n_p$ and assuming $n_p \geq 1$, we can determine the maximum purity number of a seperable *s*-particle sub-state as follows:

$$\gamma_s^{\max} = \binom{n_p}{s} = \frac{n_p!}{(n_p - s)! s!}, \quad (s < n_p). \tag{10}$$

For $0 \leq \gamma_s < \gamma_s^{\max}$ and $n_p > 1$, at least two particles are entangled. In the case of $\gamma_s = \gamma_s^{\max}$, due to mixed sub-states, entanglement can be present in the system. In other words, This condition is not sufficient for the separability of the principal state but it is necessary. It should be noted that, in general, a mixed state is not given in the decomposed form (9). Additionally, $n_p$ is an unknown number that must be obtained. The following algorithm can be used to obtain $n_p$:

- First, obtain number of all separable single-particle sub-states ($\gamma_1$).
- Perform partial tracing with respect to all separable single-particle sub-states. This step ensures that these particles do not appear in the $\gamma_k$ ($k > 1$) count. Then, the reduced density matrix $\bar{\rho}_1$ is obtained.
- Next, obtain number of all separable two-particle sub-states ($\gamma_2$) from $\bar{\rho}_1$.
- Perform partial tracing with respect to all separable two-particle sub-states from $\bar{\rho}_1$, resulting in the reduced density matrix $\bar{\rho}_{12}$.
- Continue this process untill $\gamma_{n-1}$ is obtained.

Finally, $n_p$ can be obtained from the following equation:

$$n_p = \sum_{s=1}^{n-1} s \gamma_s. \tag{11}$$

## 2.3. Entanglement rate

One question raised pertains to the entanglement rate of particles in *n*-particle pure states. To address this, the proposed method calculates the entanglement rate, initially explained through an example and then generalized for an arbitrary *n*-particle pure state. Suppose the below state:

$$|\Psi\rangle = |GHZ\rangle|\Phi^+\rangle, \tag{12}$$

where $|GHZ\rangle = 1/\sqrt{2}(|0_A 0_B 0_C\rangle + |1_A 1_B 1_C\rangle)$ and $|\Phi^+\rangle = 1/\sqrt{2}(|0_D 0_E\rangle + |1_D 1_E\rangle)$. When measuring the entanglement rate, it's inadequate to simply ratio the number of entangled particles to the total number in the system. This approach yields misleading results; for instance, in the case of $|\Psi\rangle$, where the entanglement appears maximal despite particles $A$, $B$, and $C$ not being entangled with $E$ and $D$. In this research, a more refined method is proposed: calculating the entanglement rate of a sub-state involves the ratio of coupled particles engaged in entanglement to all possible couple particles within the system. For the sub-states $|GHZ\rangle$ and $|\Phi^+\rangle$, these ratios are 0.3 and 0.1, respectively.

In general, for an entangled $s$-particle sub-state, the number of particle pairs involved in entanglement can be obtained as follows:

$$\binom{s}{2} = \frac{s(s-1)}{2}. \tag{13}$$

By replacing $s$ with $n$, we can derive the total number of possible particle pairs within the system. Subsequently, equation (13) outlines the entanglement rate for an entangled $s$-particle sub-state $|\eta\rangle$.

$$e_\eta = \frac{s(s-1)}{n(n-1)}, \tag{14}$$

where $e_\eta$ represents the entangled rate and $n$ denotes the total number of particles in the system. Then the total entanglement rate of the system reads:

$$e_{total} = \sum_i e_{\eta_i}. \tag{15}$$

Eq. (15) serves as an entanglement measure for an $n$-particle pure state, given that all its entangled sub-states are maximally entangled. Notably, for $e_{total} = 0$, the system is separable, whereas for $e_{total} = 1$, it achieves maximal entanglement. Table (1) provides the entanglement rates for various $n$-partite pure states.

| Quantum state | Entanglement rate |
|---|---|
| $|\psi_1\rangle = |\Phi^+\rangle^{\otimes 5}$ | 1/9 |
| $|\psi_2\rangle = |GHZ\rangle^{\otimes 5}$ | 1/7 |
| $|\psi_3\rangle = |GHZ\rangle_5 \otimes |GHZ\rangle_4$ | 4/9 |
| $|\psi_4\rangle = |GHZ\rangle_5 \otimes |\Phi^+\rangle^{\otimes 4}$ | 7/39 |

Table 1: Entanglement rate of several $n$-particle pure states. $|GHZ\rangle_m$ indicates a generalized m-qubit GHZ state.

## 3. Conclusion

In this paper, it was explained that the maximum purity number, introduced in Section 2, is a necessary and sufficient criterion for the separability of $n$-particle pure states and a necessary criterion for the separability of mixed states in which there is at least one two-particle pure sub-state. If the purity number is not maximum, there are at least two entangled particles in the principal state. In comparison to other criteria, which often involve complex mathematical operations or are difficult to apply to large systems, the maximum purity number stands out for its practical applicability, especially in large-scale quantum systems. While it might not capture all nuances of entanglement in mixed states (where additional criteria might be necessary), its simplicity and scalability make it a practical and effective tool for analyzing pure quantum states.

Furthermore, the concept of "entanglement rate" was introduced. The entanglement rate quantifies the proportion of particles within the system that are entangled with at least one other particle. If all entangled sub-states are maximally entangled, the entanglement rate can be considered an entanglement measure. The study's findings provide useful tools for characterizing the intricate entanglement properties of $n$-particle states, which may have important implications for quantum information processing, where the control and understanding of entanglement are essential.

**CRediT authorship contribution statement**

the author confirms sole responsibility for the following: study conception and design, data collection, analysis and interpretation of results, and manuscript preparation.

**Declaration of competing interest**

The author declare that he has no known competing financial interests or personal relationships that could have appeared to influence the work reported in this paper.

**Data Availability**

no data associated in the article.

**Acknowledgements**

This research has received no external funding.